%% file: main.tex
\documentclass{article}

\usepackage[preprint]{neurips_2024}
\usepackage{natbib}

\usepackage[utf8]{inputenc} 
\usepackage[T1]{fontenc}    
\usepackage{hyperref}       
\usepackage{url}            
\usepackage{booktabs}       
\usepackage{amsfonts}       
\usepackage{nicefrac}       
\usepackage{microtype}      
\usepackage{xcolor}         
\usepackage{graphicx}
\usepackage{subcaption}
\usepackage{wrapfig}
\usepackage{todonotes}
\usepackage{array}
\usepackage{booktabs}
\usepackage{amssymb}
\usepackage{pifont}
\usepackage{tablefootnote}
\usepackage{tabularx}

\newcommand{\sysname}{AIBrix}
\renewcommand{\arraystretch}{1.3} 

\title{\sysname{}: Towards Scalable, Cost-Effective Large Language Model Inference Infrastructure}

\author{%
  The AIBrix Team\\
  \texttt{maintainers@aibrix.ai} \\
}

\begin{document}

\maketitle

\begin{abstract}
  We introduce \sysname{}, a cloud-native, open-source framework designed to optimize and simplify large-scale LLM deployment in cloud environments.
  Unlike traditional cloud-native stacks, \sysname{} follows a co-design philosophy, ensuring every layer of the infrastructure is purpose-built for seamless integration with inference engines like vLLM.

  \sysname{} introduces several key innovations to reduce inference costs and enhance performance including high-density LoRA management for dynamic adapter scheduling, LLM-specific autoscalers, and prefix-aware, load-aware routing.
  To further improve efficiency, \sysname{} incorporates a distributed KV cache, boosting token reuse across nodes, leading to a 50\% increase in throughput and a 70\% reduction in inference latency.
  \sysname{} also supports unified AI runtime which streamlines model management while maintaining vendor-agnostic engine compatibility.

  For large-scale multi-node inference, \sysname{} employs hybrid orchestration—leveraging Kubernetes for coarse-grained scheduling and Ray for fine-grained execution—to balance efficiency and flexibility.
  Additionally, an SLO-driven GPU optimizer dynamically adjusts resource allocations, optimizing heterogeneous serving to maximize cost efficiency while maintaining service guarantees.
  Finally, \sysname{} enhances system reliability with AI accelerator diagnostic tools, enabling automated failure detection and mock-up testing to improve fault resilience.
  \sysname{} is available at \textcolor{red}{https://github.com/vllm-project/aibrix}.

\end{abstract}

\input{text/intro}

\input{text/motivation}
\input{text/architecture}
\input{text/conclusion}



\bibliographystyle{plainnat}
\bibliography{main.bib}



\appendix
\section{List of Authors}

\sysname{} was made possible by the following team members:

\noindent
\renewcommand{\arraystretch}{1.5} 
\begin{tabularx}{\textwidth}{@{}l>{\hspace{1.5cm}}X X r@{}}  
    Jiaxin Shan\textsuperscript{1} & Varun Gupta\textsuperscript{1} & Le Xu\textsuperscript{1} & Haiyang Shi\textsuperscript{1} \\
    Jingyuan Zhang\textsuperscript{1} & Ning Wang\textsuperscript{1} & Linhui Xu\textsuperscript{1} & Rong Kang\textsuperscript{1} \\
    Tongping Liu\textsuperscript{1} & Yifei Zhang\textsuperscript{1} & Yiqing Zhu\textsuperscript{1} & Shuowei Jin\textsuperscript{1,2} \\
    Gangmuk Lim\textsuperscript{1,3} & Binbin Chen\textsuperscript{1} & Zuzhi Chen\textsuperscript{1} & Xiao Liu\textsuperscript{1} \\
    Xin Chen\textsuperscript{1} & Kante Yin\textsuperscript{4} & Chak-Pong Chung\textsuperscript{1} & Chenyu Jiang\textsuperscript{1} \\
    Yicheng Lu\textsuperscript{1} & Jianjun Chen\textsuperscript{1} & Caixue Lin\textsuperscript{1} & Wu Xiang\textsuperscript{1} \\
    Rui Shi\textsuperscript{1} & Liguang Xie\textsuperscript{1} & & \\
\end{tabularx}

\footnotetext[1]{ByteDance.}
\footnotetext[2]{University of Michigan. Work done as part of the internship. }
\footnotetext[3]{University of Illinois at Urbana Champaign. Work done as part of the internship.}
\footnotetext[4]{DaoCloud}

Each contributor played a vital role in ensuring the success of AIBrix through technical development, performance evaluation, and strategic business guidance.



\end{document}

%% file: text/intro.tex
\section{Introduction}
\label{sec:intro}
 
Large language models (LLMs) have revolutionized AI applications, powering innovations in areas like chatbots, automated content generation, and advanced recommendation engines.
While API services based on proprietary models like OpenAI and Anthropic have gained widespread adoption, many enterprises often seek open-source alternatives due to data security concerns, customizability, or the cost of of proprietary solutions.
The growing demand for hosting open-source models like LLaMA, Deepseek, Qwen and Mistral and offer production-grade APIs presents new challenges -- namely, how to efficiently deploy them at scale while maintaining low inference latency and cost efficiency.


Deploying LLMs in production requires more than just an optimized model; it demands a holistic system approach spanning multiple layers:

\begin{itemize}
    \item \textbf{Open-Source Models}: The foundation of AI applications, where model optimizations such as model architecture, Multi-Head Latent Attention (MLA) (\cite{liu2024deepseek}), distillation, and adapter fine-tuning enhance performance and adaptability.

    \item \textbf{Inference Engines}: Inference engines like vLLM(\cite{kwon2023efficient}) or TensorRT-LLM improve model serving efficiency via KV cache management, model parallelism, attention optimization, and other optimizations.

    \item \textbf{System-Level Orchestration (\sysname{})}: The critical but often overlooked layer that determines real-world cost efficiency and scalability by governing resource scheduling, autoscaling, request routing, heterogeneity management, and multi-cluster or multi-region resource optimization.
\end{itemize}

While model and engine optimizations are crucial, system-level orchestration is the key to unblocking true cost efficiency.
Without a well-designed infrastructure, even the most advanced model and inference engines struggle with real-world deployment challenges, such as autoscaling, cache-aware routing, heterogeneity resource management. 

To address these challenges, we introduce \sysname{}, a novel cloud-native framework designed to simplify and optimize LLM inference infrastructure, providing users with a \textbf{one-click deployment experience} while ensuring best-in-class performance and cost-efficiency.
Our key contributions include:

\begin{itemize}
    \item \textbf{High-Density LoRA Management:} \sysname{} 
    enables dynamic LoRA registration and lineage support (\cite{loraproposal}) in vLLM, streamlining LoRA adapter management and reducing the cost of managing fine-tuned models.

    \item \textbf{LLM-Specific Autoscaling:} \sysname{} supports various scenario-driven LLM autoscaling policies. It also features optimizations such as sliding window metric aggregation to reduce the propagation delay of real-time metrics. 
    
    \item \textbf{Advanced LLM Gateway and Routing Strategies:} \sysname{} introduces an LLM-aware API gateway, extending Envoy Gateway to optimize instance routing and support various routing strategies.
    Unlike traditional gateways that distribute requests blindly, \sysname{} analyzes token patterns, prefill cache availability, and compute overhead to enhance routing efficiency in diverse deployment scenarios. 
    
    \item \textbf{Unified AI Runtime with GPU Streaming Loader:} \sysname{} serves as a unified runtime layer, managing interactions between inference engine pods and the control plane. It automates models artifact handling, configures inference engines, and provides real-time observability, ensuring vendor-agnostic compatibility.
    Additionally, \sysname{} features a GPU streaming loader that bypasses disk I/O bottlenecks to accelerate model loading and execution.
        
    \item \textbf{Distributed and disaggregated KV Cache pool:} \sysname{} introduces a distributed KV cache that enables high-capacity, cross-engine KV reuse while optimizing network and memory efficiency. 
    Key innovations include a scan-resistant eviction policy, reduced redundant data transfers, asynchronous metadata updates, and shared-memory-based data exchange, enhancing inference throughput and efficiency.
    
    \item \textbf{Mixed-Grain Multi-Node Inference Orchestration:} \sysname{} introduces a hybrid approach to multi-node inference by integrating Ray (\cite{ray}) for fine-grained application orchestration with Kubernetes for coarse-grained resource management.
    Compared to inference engine's native supports in a distributed environment (\cite{distributed-serving}), which emphasize parallelism over service-oriented needs, AIBrix balances distributed execution with production-grade orchestration, achieving scalability, rolling upgrades, and efficient resource allocation. 
    
    \item \textbf{Cost efficient and SLO-driven Heterogeneous Serving:} \sysname{} introduces a GPU optimizer that balances cost efficiency with SLO adherence, dynamically selecting the optimal GPU configuration based on workload characteristics and hardware availability, ensuring cost-effective heterogeneous GPU utilization.
    
    \item \textbf{Accelerator Diagnostic and Failure Mockup Tools:} \sysname{} introduces a diagnostic tool that leverages AI accelerators' built-in capabilities to detect and diagnose hardware failures.
    Also, a failure mockup tool simulates hardware failures, enabling rigorous fault tolerance and recovery testing.
    
\end{itemize}

\sysname{} is a cloud-native, open-source framework that simplifies and optimizes LLM deployment in production environments, offering AI practitioners across industry and academia with a flexible, scalable, and cost-effective serving solution.
By deeply integrating model and engine optimizations with system-level orchestration, \sysname{} bridges the gap between efficiency and flexibility, setting a new standard for large-scale LLM inference workloads.



%% file: text/motivation.tex
\section{Related Works}
\label{sec:motivation}

Existing cloud-native and machine learning (ML) serving frameworks provide fundamental infrastructure for model inference but lack key optimizations necessary for large-scale LLM inference. We categorize prior works into two main areas: microservice-based serverless frameworks and traditional ML model-serving frameworks, highlighting their limitations when applied to LLM workloads.

\textbf{Microservice-Based Systems}
Microservice-based frameworks like Knative (\cite{knative}) and Istio (\cite{istio}) offer powerful solutions for managing stateless services, emphasizing advanced traffic control mechanisms such as request rate limiting, request interception, authentication (AuthN), authorization (AuthZ), and autoscaling based on QPS and concurrency metrics. However, these solutions are not designed for GPU-based inference workloads and fail to address the fundamental changes that LLM applications introduce. Unlike traditional microservices, LLM inference does not require complex service meshes or extensive request routing features. Instead, it introduces new challenges such as token-based rate limiting, KV cache-aware autoscaling, and model-specific scheduling constraints. For example, circuit-breaker-based rate limiting in Knative is incompatible with LLM’s token-based inference constraints, and QPS-based autoscaling cannot accurately capture GPU-bound resource usage patterns such as KV cache memory pressure. Furthermore, the overhead of Istio’s service mesh makes it unsuitable for LLM applications, where lightweight, inference-specific optimizations are more effective.

\textbf{Traditional ML Model-Serving Systems}
Traditional ML model-serving frameworks like KServe (\cite{kserve}) and RayServe (\cite{rayserve}) provide solutions for model deployment, request handling, and autoscaling, making them well-suited for conventional deep learning inference. These frameworks support features such as model URI management, dynamic scaling, and model versioning. However, they lack deep integration with LLM inference engines and fail to address key challenges specific to LLM workloads. LLM inference introduces unique characteristics such as highly variable input-output lengths, massive model sizes, and stateful execution (e.g., KV cache management), which require custom routing, model distribution strategies, and GPU-aware scheduling. While KServe and RayServe can deploy LLM models, they do not provide specialized optimizations for efficient batch scheduling, KV cache coordination, or heterogeneous GPU utilization. As a result, they cannot fully leverage the performance potential of modern LLM inference engines such as vLLM and TensorRT-LLM (\cite{tensorrtllm}).

%% file: text/architecture.tex
\section{\sysname{}: Cloud-native Infrastructure for LLM-serving}
\label{sec:architecture}

\subsection{\sysname{} Architecture}

\begin{figure}[ht]
    \centering
    \includegraphics[width=1\textwidth]{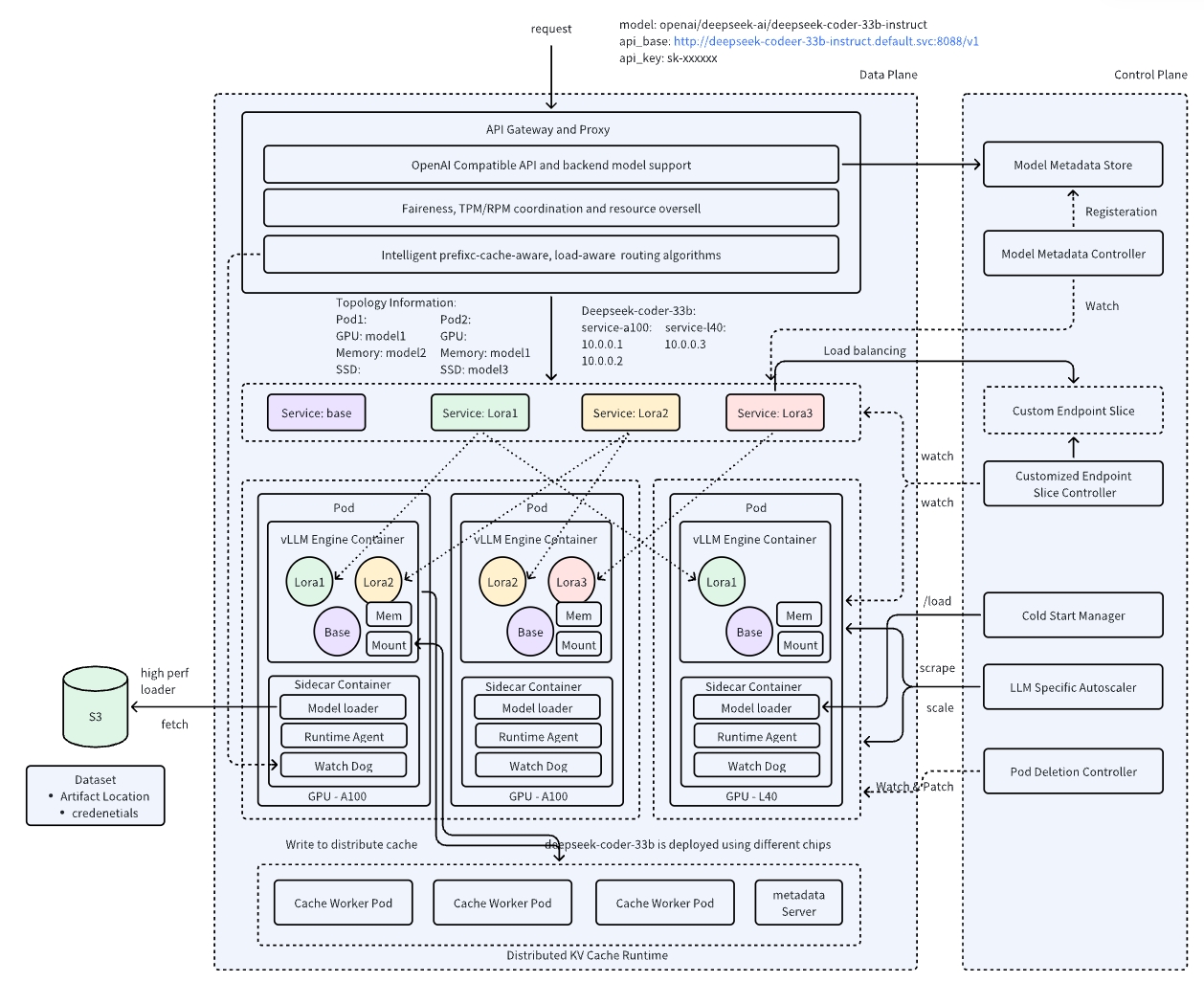}
    \caption{\textit{\sysname{} Architecture Overview.}}
    \label{fig:architecture}
\end{figure}

In this section, we describe the architecture of \sysname{}, as shown in Figure~\ref{fig:architecture}. \sysname{} contains both control plane components and data plane components. The components of the control plane manage the registration of model metadata, autoscaling, model adapter registration, and enforce various types of policies. Data plane components provide configurable components for dispatching, scheduling, and serving inference requests, enabling flexible and high-performance model execution.

\paragraph{The \sysname{} Control Plane.} The \sysname{} control plane streamlines LLM deployment by automating model management, optimizing resource allocation, and enabling intelligent scaling. AIbrix at this moment does not build any abstractions for base models, it still leverage kubernetes deployment for basic lifecycle operations. The LoRA Adapter controller enhances flexibility by enabling multi-LoRA-per-pod deployments, improving scalability and resource utilization. Another controller named RayClusterFleet manages multi-node inference for large scale models served by vLLM.

To bridge inference engines with control plane, \sysname{} features an AI Runtime that acts as a lightweight sidecar, offloading management tasks, enforcing policies, and abstracting engine interactions for systems like vLLM, SGLang (\cite{zheng2024sglang}), and TensorRT-LLM. Complementing this, the Cold Start Manager tracks model artifacts across DRAM, local storage, and cloud storage, ensuring models are loaded on the fastest available node to minimize startup latency. 

Meanwhile, the LLM-Specific Autoscaler enables real-time, millisecond-level scaling, leveraging KV cache utilization and inference-aware metrics to optimize resource allocation dynamically. Together, these components form a highly adaptive control plane, integrating cloud infrastructure and inference engines co-design to deliver scalable, cost-efficient, and high-performance LLM inference.

\paragraph{The \sysname{} Data Plane.} \sysname{} features a data plane that is responsible for handling user requests, managing model instances, executing inference, and optimizing caching strategies to ensure efficient and scalable LLM serving. It is designed to be both application-aware and resource-aware, integrating deep inference optimizations with cloud-native orchestration.

At the entry point, the API Gateway serves as the central request dispatcher, enforcing fairness policies, rate control (TPM/RPM), and workload isolation, while dynamically optimizing traffic based on KV cache locality, token throughput, and GPU heterogeneity. This adaptive routing mechanism ensures efficient inference execution across diverse hardware configurations.

Model execution is facilitated by the Serving Unit, which consists of both the Inference Engine and the AI Runtime sidecar. \sysname{} also introduces a distributed KV Cache Runtime, which extends external cache services to manage dynamically generated KV cache during inference. KV cache reuse is crucial for reducing redundant computation and improving token generation efficiency. The distributed DRAM-based KV cache runtime enables scalable, low-latency cache access across nodes and supports advanced optimizations, such as prefix cache expansion, future prefill/decode disaggregation remote pool (\cite{zhong2024distserve}) and request migration, further improving performance in memory-constrained environments and play a crucial role when certain features are incompatible. For example, in vLLM v0.7.1, when MLA is enabled for Deepseek-R1, prefix cache must be disabled in vLLM.

Together, these components form a highly optimized data plane, ensuring that \sysname{} delivers scalable, cost-effective, and high-performance inference for LLM applications while balancing latency, resource utilization, and workload fairness.

\subsection{\sysname{} Features}

\sysname{} brings together a few innovations to streamline enterprise-grade LLM infrastructure, enhancing scalability and efficiency. 

\subsubsection{High-Density LoRA Management.} 

Scaling Low-Rank Adaptation (LoRA) (\cite{hulora}) fine-tuned models has traditionally been constrained by rigid deployments, limiting flexibility and increasing costs. Conventional serving infrastructures treat LoRA adapters as static attachments to a base model, making dynamic scaling impractical. The lack of integrated resource management results in inefficient allocation, unreliable evictions, and suboptimal failure handling.

\sysname{} introduces high-density LoRA management (Figure~\ref{fig:lora}), enabling dynamic adapter loading and unloading, intelligent scheduling, and LoRA-aware routing to enhance inference efficiency. By dynamically registering LoRA adapters, \sysname{} supports high-density deployments, significantly reducing inference costs—particularly beneficial for long-tail scenarios. Leveraging Kubernetes' Service and EndpointSlice mechanisms, \sysname{} optimizes LoRA model discovery and placement, minimizing interference and maximizing resource utilization. Additionally, enhancements to vLLM strengthen LoRA management capabilities, reducing operational overhead and improving inference performance under mixed workloads.

\begin{minipage}{0.48\textwidth}
    \centering
    \includegraphics[height=.7\textwidth]{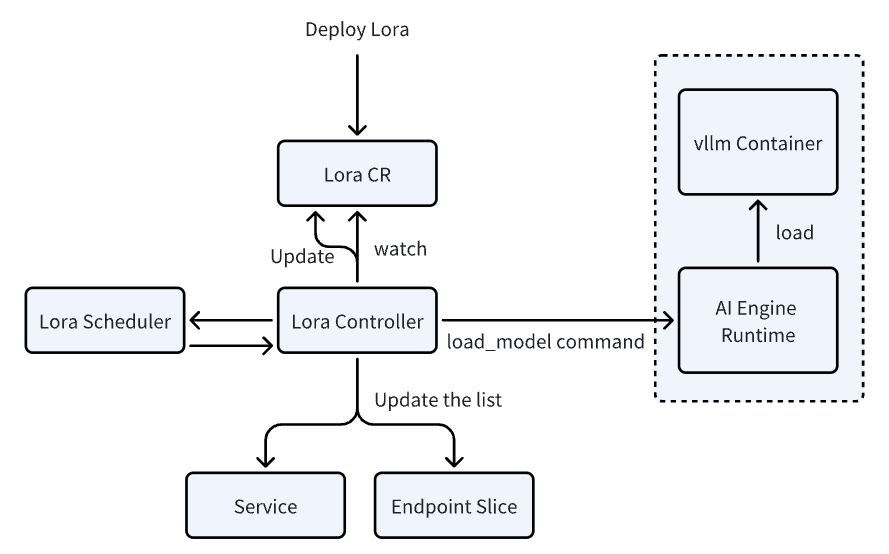}
    \captionof{figure}{\textit{High-Density LoRA Management.}}
    \label{fig:lora}
\end{minipage}
\hfill
\begin{minipage}{0.48\textwidth}
    \centering
    \includegraphics[height=.7\textwidth]{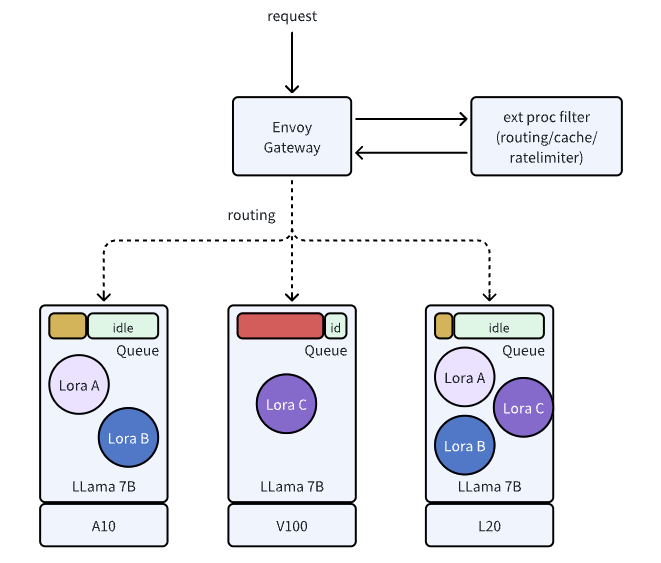}
    \captionof{figure}{\textit{Advanced LLM Gateway and Routing.}}
    \label{fig:routing}
\end{minipage}

\subsubsection{Advanced LLM Gateway and Routing Strategies} 

Traditional API gateways struggle with LLM inference due to the diverse complexities of requests, ranging from simple queries to multi-turn interactions that require intricate token management. Generic routing leads to inefficient traffic distribution and latency spikes. \sysname{} addresses this by providing an LLM-aware gateway, extending the Envoy Gateway (\cite{envoy}) to support instance routing, prefix cache awareness, and least-GPU-memory-based strategies, as shown in Figure~\ref{fig:routing}. Unlike conventional systems that blindly distribute requests, \sysname{} analyzes token patterns, prefill cache availability, and compute overhead to optimize traffic flow. This enables advanced routing customization and user-defined strategies. For each pending request, the current version of \sysname{} determines the target instance based on one of the following routing policies: 
\begin{itemize} 
    \item \texttt{random}: Randomly selects an available instance.
    \item \texttt{throughput}: Selects the instance with the lowest throughput in terms of tokens per second.
    \item \texttt{least-request}: Selects the instance with the lowest number of admitted requests.
    \item \texttt{least-kv-cache}: Selects the instance with the lowest average KV cache usage. 
    \item \texttt{least-latency}: Selects the instance with the lowest average request latency. For each request, this is derived from the sum of its queuing latency and serving latency.
    \item \texttt{prefix-cache-aware}: Prioritizes instances with reusable prefix cache, with a cache hit exceeding the threshold.
\end{itemize}
By selecting a fitting routing strategy, \sysname{} is able to reduce mean latency by 19.2\% and P99 latency by 79\%, ensuring efficient and fair LLM inference at scale. AIBrix team is also working with Google and other contributors on gateway-api-inference-extension \cite{gateway-api-inference-extension} project for future adoption.

\subsubsection{Unified AI Runtime with GPU Streaming Loader} 

The domain of inference engines is evolving rapidly, with certain engines quickly surpassing others in terms of performance. Additionally, some engines offer unique features, such as specialized support for specific model optimizations or feature combinations. As a result, users often prefer to leverage different engines based on their performance benefits or feature sets.

\begin{figure}[ht]
    \centering
    \includegraphics[width=.5\textwidth]{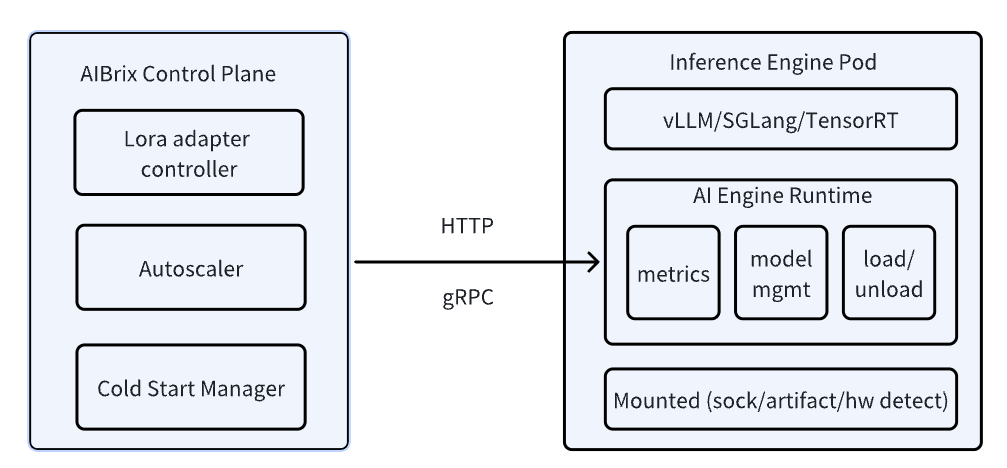}
    \caption{\textit{\sysname{}'s unified AI runtime}}
    \label{fig:runtime}
\end{figure}

However, directly supporting these engines in the control plane is not scalable due to the wide variety of protocols they use. To address this, an abstraction layer is necessary to unify and streamline interactions with these diverse inference engines, allowing for seamless integration and more efficient management. \sysname{} introduces a unified AI runtime (Figure~\ref{fig:runtime}), acting as a bridge between the \sysname{} Control Plane and inference engine pods. This runtime manages models, configures engines, provides observability, and enables vendor-agnostic support. It ensures seamless communication between core components, such as the LoRA adapter controller, autoscaler, and cold start manager, facilitating dynamic, cloud-native resource management.

\subsubsection{LLM-Specific Autoscaling for Performance Optimization} 
Autoscaling for LLM inference presents unique challenges due to DCGM (\cite{dcgm}) metric limitations, non-linear scaling behaviors, and the inadequacy of traditional indicators like QPS or concurrency. Request complexity and I/O size vary widely, often overwhelming systems before autoscalers can react. Moreover, large model images and slow model distribution introduce a 2-3 minute delay for new pods, making rapid scaling inefficient. \sysname{} mitigates these challenges by supporting configurable LLM-specific autoscaling policies. It bypasses the custom metrics path and maintains sliding window metric aggregation directly in the autoscaler for real-time load reporting. By leveraging advanced autoscaling algorithms such as Knative Pod Autoscaler (KPA) (\cite{kpa}) and \sysname{} Pod Autoscaler (APA) (\cite{huo2023high}), \sysname{} is able to reduce latency by 11.5\%, increases token throughput by 11.4\%, and minimizes scaling oscillations by 33\% compared to native HPA. Future work explores token-based proactive scaling and SLO-driven autoscaling for enhanced efficiency and responsiveness.

\subsubsection{Distributed KV Cache Pool}

The growing demand for large language models necessitates efficient memory management and caching strategies to optimize inference performance and reduce costs. In multi-turn applications such as chatbots and agent-based systems, overlapping token sequences often lead to redundant computations during the prefill phase, resulting in resource wastage and limited throughput. While inference engines like vLLM incorporate built-in KV caching, single-node caches suffer from memory constraints, engine-specific storage limitations, and a lack of support for KV migration and prefill-decode disaggregation.

\begin{figure}[ht]
    \centering
    \includegraphics[width=1\textwidth]{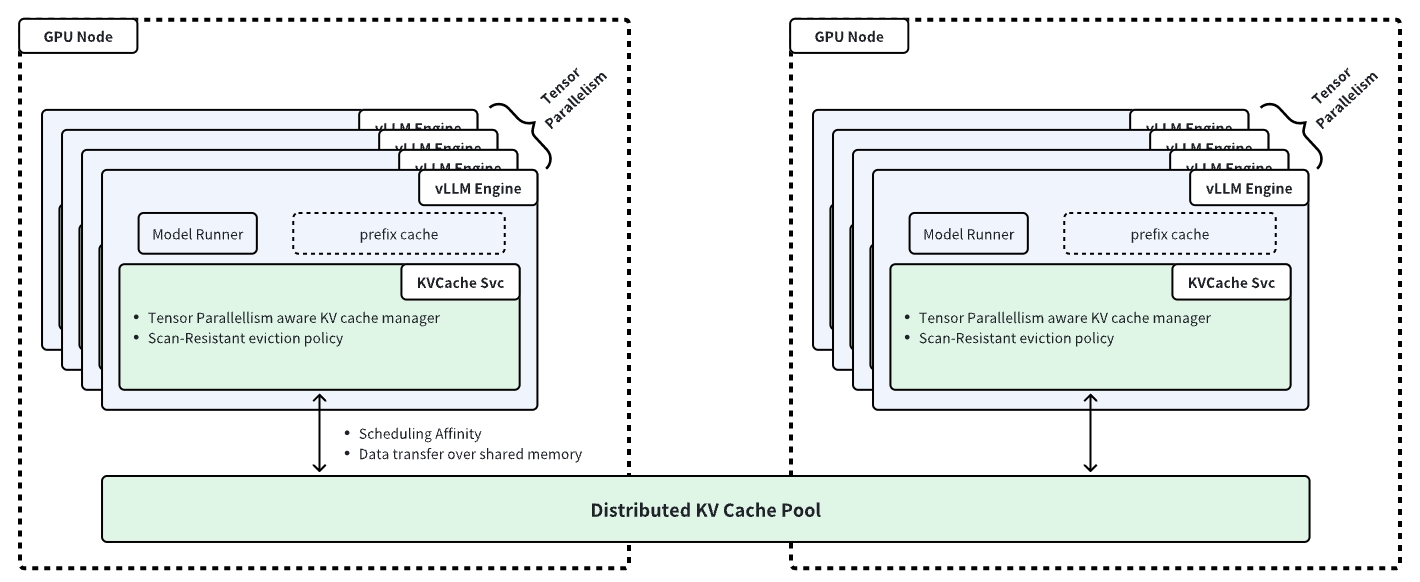}
    \caption{\textit{Distributed KV Cache Pool.}}
    \label{fig:cachepool}
\end{figure}

To address these challenges, \sysname{} introduces a distributed KV cache (Figure~\ref{fig:cachepool}), enabling high-capacity, cross-engine KV reuse while optimizing network and memory efficiency. The system employs a scan-resistant eviction policy to selectively persist hot KV tensors, reducing unnecessary data transfers. Additionally, asynchronous metadata updates minimize overhead, while cache-engine colocation accelerates data transfer through shared memory.

Table~\ref{tab:eval_cache_performance_comparison} presents the performance evaluation of our distributed KV cache using the Bird-SQL benchmark (\cite{birdsql}), conducted on 4 $\times$ Nvidia A10 GPUs. Our results indicate that, even when compared to vLLM’s built-in prefix caching, combining the distributed KV cache with prefix caching improves peak throughput by approximately 50\%, reduces average and P99 TTFT by approximately 60\% and 70\%, respectively, and lowers average and P99 ITL by approximately 30\% and 70\%, respectively. These findings demonstrate significant efficiency gains.

\begin{table}[ht]
    \centering
    \scriptsize
    \begin{tabular}{>{\raggedright\arraybackslash}p{2cm} ccccccccc}
        \toprule
        & \multicolumn{2}{c}{Tokens} & \multicolumn{2}{c}{Throughput (tokens/sec)} & \multicolumn{2}{c}{TTFT (ms)} & \multicolumn{2}{c}{ITL (ms)} & \multicolumn{1}{c}{Completion} \\
        \cmidrule(r){2-3} \cmidrule(r){4-5} \cmidrule(r){6-7} \cmidrule(r){8-9}
        Method & Prompt & Decoding & Total & Decoding & Avg. & P99 & Avg. & P99 & Time (sec) \\
        \midrule
        vLLM Default & 1082837 & 12726 & 1,802.30 & 20.94 & 3,067.07 & 10,060.29 & 189.04 & 3,175.79 & 607.87 \\
        AIBrix Distributed KV Cache + Default & 1082837 & 12762 & 4,133.45 & 48.15 & 825.77 & 2,132.97 & 89.78 & 831.98 & 265.06 \\
        \textbf{Improvement} & & & 129.34\% & 129.98\% & 73.08\% & 78.80\% & 52.51\% & 73.80\% & 56.40\% \\
        \midrule
        vLLM Chunked Prefill & 1082837 & 12756 & 1,820.63 & 21.20 & 4,500.58 & 12,411.33 & 164.41 & 288.77 & 601.77 \\
        AIBrix Distributed KV Cache + Chunked Prefill & 1082837 & 12744 & 3,320.91 & 38.63 & 2,235.12 & 6,151.27 & 93.34 & 151.68 & 329.90 \\
        \textbf{Improvement} & & & 82.40\% & 82.23\% & 50.34\% & 50.44\% & 43.23\% & 47.47\% & 45.18\% \\
        \midrule
        vLLM Prefix Caching & 1082837 & 12775 & 3,703.26 & 43.18 & 999.95 & 5,744.27 & 99.50 & 1,653.56 & 295.85 \\
        AIBrix Distributed KV Cache + Prefix Caching & 1082837 & 12761 & 5,615.71 & 65.41 & 349.39 & 1,306.05 & 69.95 & 456.31 & 195.10 \\
        \textbf{Improvement} & & & 51.64\% & 51.48\% & 65.06\% & 77.26\% & 29.70\% & 72.40\% & 34.06\% \\
        \bottomrule
    \end{tabular}
    \caption{\textit{Performance comparison of vLLM and \sysname{} Distributed KV Cache with different configurations. Here the vLLM Default is the configuration with chunked prefill and prefix caching disabled.}}
    \label{tab:eval_cache_performance_comparison}
\end{table}

\subsubsection{Mix-Grain Multi-Node Inference Orchestration}
The release of Llama-3.1-405B (\cite{llama3-405b}) and Deepseek-R1 (\cite{deepseek-r1}) has significantly increased the demand for multi-node inference. However, existing frameworks such as vLLM prioritize parallelism over service-oriented requirements like scaling and rolling upgrades, necessitating external orchestration.

While Kubernetes and Ray provide orchestration capabilities, they come with trade-offs: Kubernetes operators offer coarse-grained resource management but can be complex for fine-grained scheduling, whereas Ray excels in distributed communication but lacks holistic resource control.

To address these limitations, \sysname{} introduces a hybrid approach that integrates Ray for fine-grained application orchestration with Kubernetes for coarse-grained resource management. This method simplifies operator design, ensuring adaptability to future paradigm shifts in inference orchestration. We observed that orchestration often changes due to inadequate techniques, such as prefill and decode (P\&D) disaggregation. Maintaining flexibility in orchestration design is crucial. Informed by Bytedance's extensive experience in managing Kubernetes and Ray workloads, \sysname{} leverages adaptive orchestration strategies to overcome workload communication challenges and optimize multi-node inference execution.

\begin{figure}[ht]
    \centering
    \includegraphics[width=.8\textwidth]{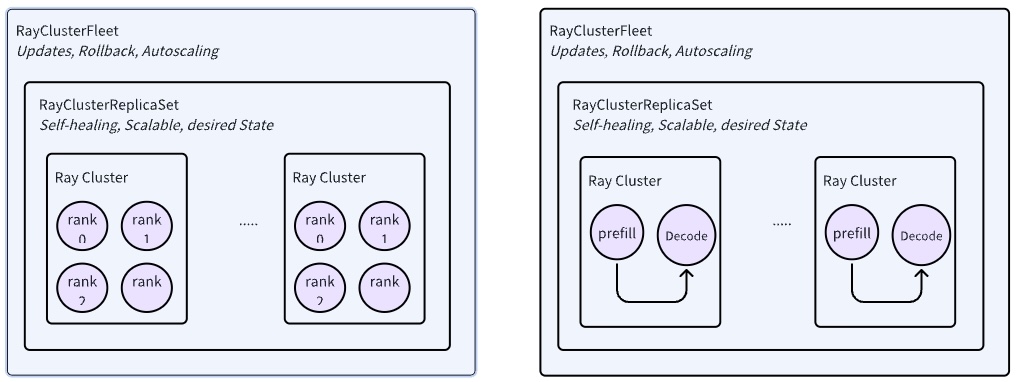}
    \caption{\textit{Mix-Grain Multi-Node Inference Orchestration.}}
    \label{fig:mixgrain}
\end{figure}

\subsubsection{Cost-Efficient and SLO-Driven Heterogeneous Serving} 

Recent research, such as M\'elange (\cite{griggs2024m}) and QLM (\cite{patke2024queue}), has demonstrated that LLM throughput under specific SLO constraints depends on input/output token counts and model selection within heterogeneous GPU environments.

Additionally, request cost-efficiency varies across GPUs even for identical models, as different GPUs exhibit distinct performance characteristics under varying workloads. Compounding these challenges, production users often face GPU availability constraints, limiting access to consistent GPU types.

\begin{figure}[ht]
    \centering
    \includegraphics[width=1\textwidth]{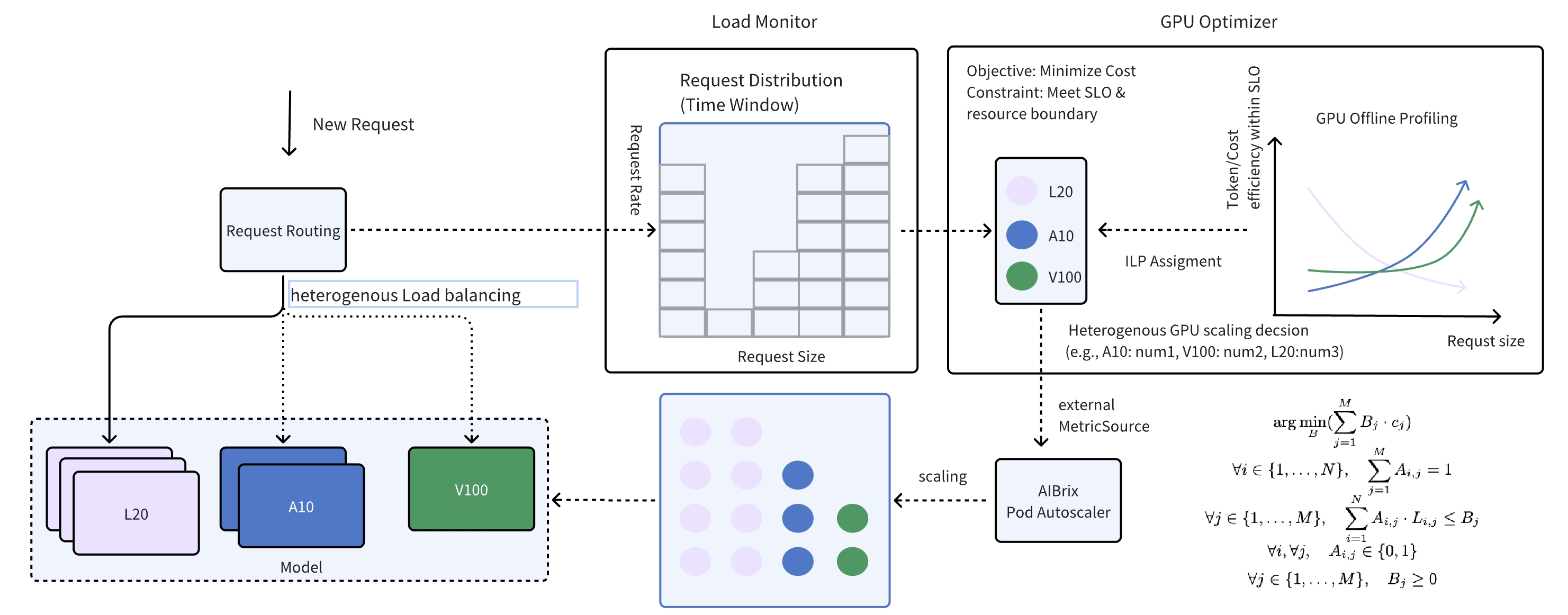}
    \caption{\textit{Cost-Efficient and SLO-Driven Heterogeneous Serving.}}
    \label{fig:hetero}
\end{figure}

To address these issues, \sysname{} introduces a GPU optimizer—an off-path component designed to optimize heterogeneous GPU serving by balancing cost efficiency and SLO adherence. As shown in Figure~\ref{fig:hetero}, the architecture consists of three key components:

\begin{figure}[ht]%
    \centering
    \subfloat[\centering \textit{Throughputs of workload using the deepseek-coder-7b model on L20, V100, and A10 GPU.}]{{\includegraphics[width=.45\columnwidth]{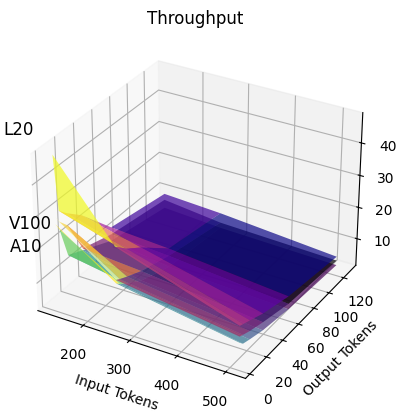}}\label{fig:hetero-throughput}}
    \hfill
    \subfloat[\centering \textit{Most requests favor L20 for cost-efficiency, while those with <200 input and <100 output tokens prefer A10.}]{{\includegraphics[width=.45\columnwidth]{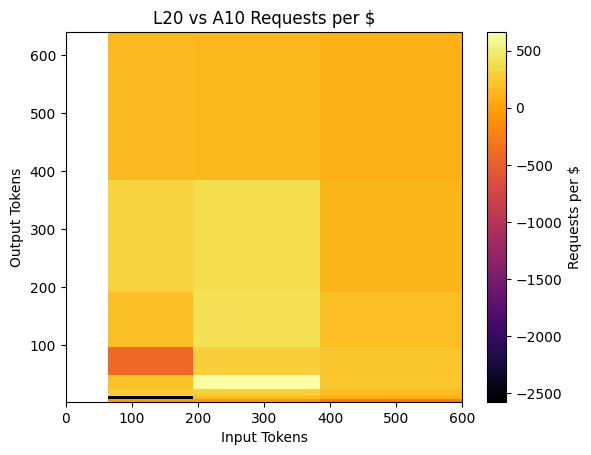}}\label{fig:hetero-cost}}
    \caption{\textit{Selection of accelerator types differ by workloads used.}}
    \label{fig:hetero-accel-types}
\end{figure}

\begin{itemize}
    \item \textbf{Load Monitor} tracks deployment changes, assumes different model deployments use distinct GPUs, and analyzes \sysname{} Gateway statistics to identify dominant workload patterns.
    
    \item \textbf{GPU Optimizer} dynamically selects the optimal GPU combination to balance cost efficiency and SLO adherence.
    
    \item \textbf{Pod Autoscaler} reads external MetricSource data from the GPU Optimizer to adjust GPU allocation dynamically. Currently, the GPU optimizer supports an ILP-based solution inspired by Melange, requiring pre-deployment profiling. \sysname{} provides toolkits for workload benchmarking and profiling.
\end{itemize}

In our experiment comparing heterogeneous workloads (using A10 (\cite{a10}) and L20 (\cite{l20})) against a homogeneous setup (using only L20), we evaluated a mixed dataset comprising ShareGPT (\cite{sharegpt}) and internal Text2SQL workloads. The heterogeneous configuration resulted in a latency increase of up to 20\% while remaining within the specified SLO. However, this setup achieved a cost reduction of approximately 10\% compared to the homogeneous GPU deployment.

\subsubsection{AI Accelerator Diagnostic and Failure Mockup Tools} 

GPU failures and performance degradation pose significant challenges in large-scale AI deployments. Silent errors, overheating, memory leaks, and intermittent failures can degrade model performance, increase latency, or even cause system crashes. Diagnosing GPU issues is particularly difficult in heterogeneous environments, where different GPU models exhibit inconsistent behavior under varying workloads.

\begin{figure}[ht]%
    \centering
    \subfloat[\centering \textit{Failure diagnosis.}]{{\includegraphics[width=.5\columnwidth]{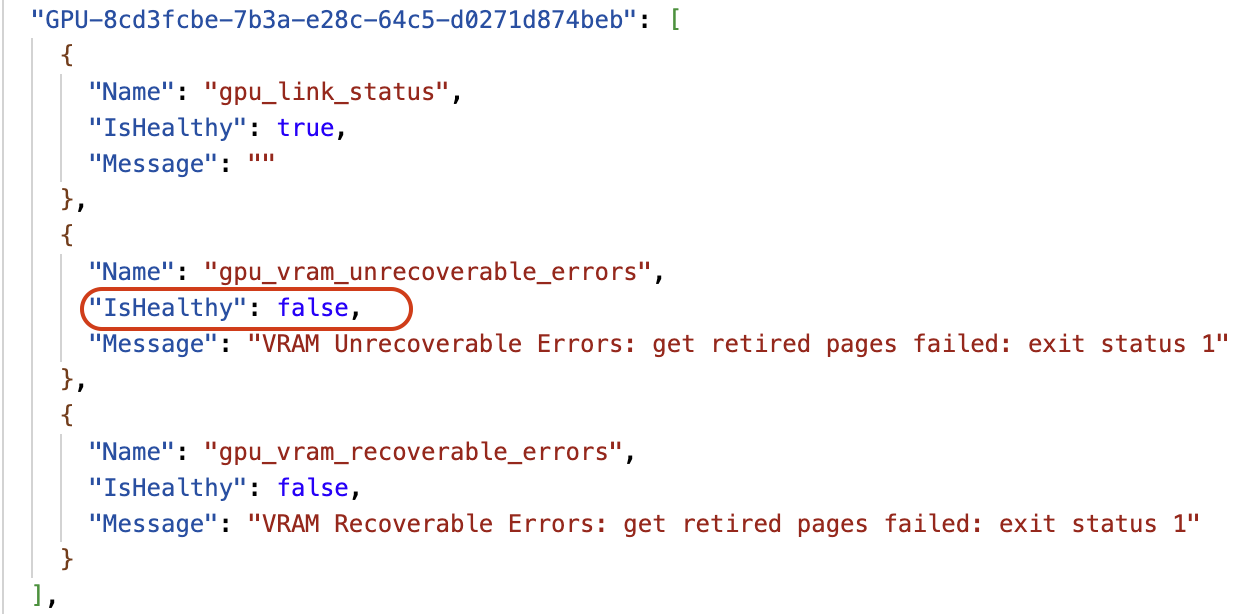}}\label{fig:diagnosis}}
    \hfill
    \subfloat[\centering \textit{Mock file.}]{{\includegraphics[width=.7\columnwidth]{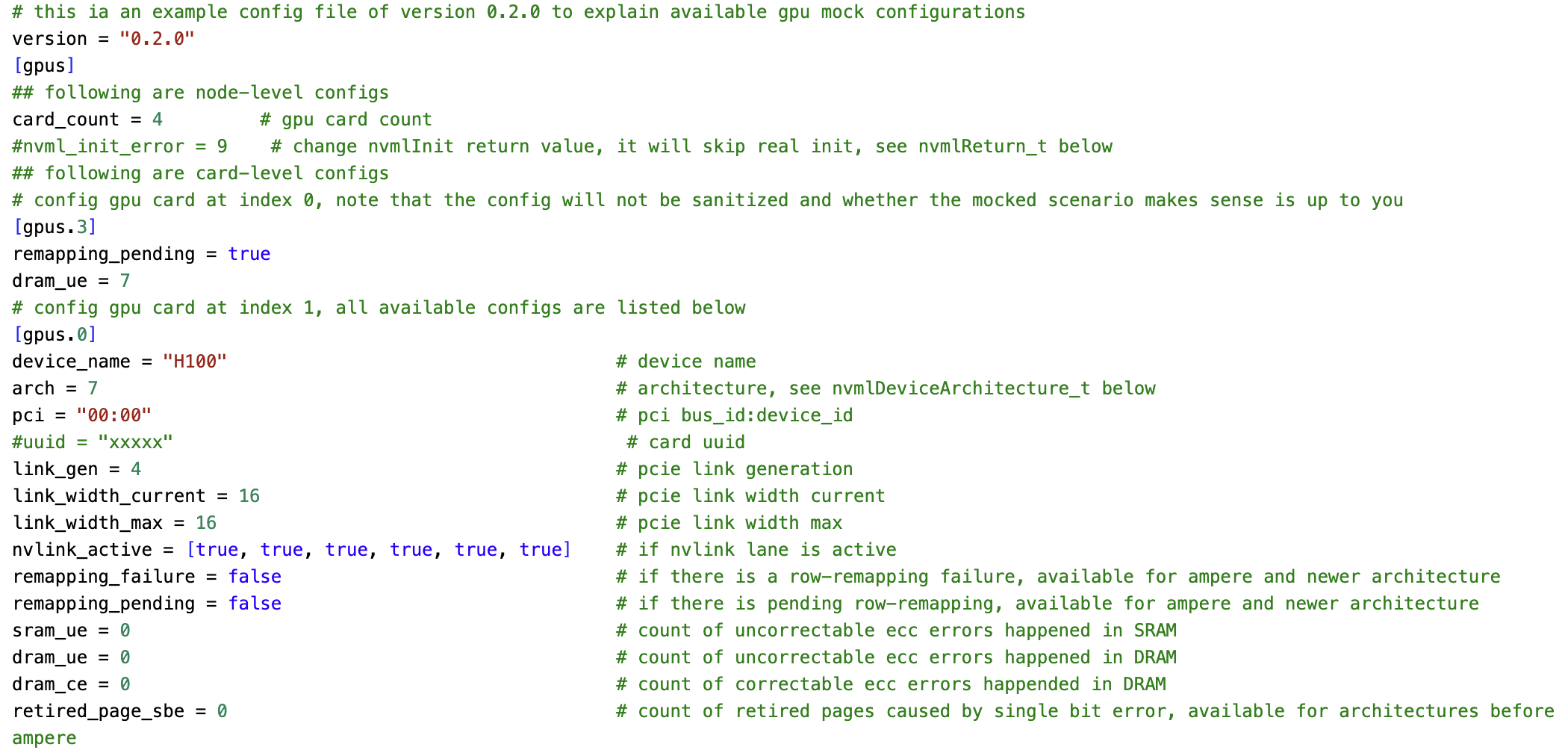}}\label{fig:mockfile}}
    \caption{\textit{\sysname{} failure diagnosis and mockup tools.}}
    \label{fig:hetero-accel-types}
\end{figure}

To tackle these challenges, AIBrix Accelerator Tools introduces:
\begin{itemize}
    \item \textbf{GPU diagnostics and issue identification.} \sysname{} automates fault detection, helping users proactively identify and resolve GPU-related performance issues before they impact workloads (Figure~\ref{fig:diagnosis}).
    \item \textbf{GPU failure mock-up tools.} \sysname{} simulates GPU failures, allowing developers to test and build resilient AI frameworks capable of recovering gracefully from hardware failures. Currently, both Nvidia GPUs and Ascend 910B NPUs are supported, with plans to extend compatibility to additional accelerators in the near future (Figure~\ref{fig:mockfile}).
\end{itemize}

%% file: text/conclusion.tex
\section{Conclusion}
\label{sec:conclusion}

\textbf{Summary.}  
We introduced \sysname{}, a novel framework designed to address the challenges of large-scale LLM inference. \sysname{} leverages serverless features, including LLM specific autoscaling, cold start optimization, and high-density deployment, to significantly reduce inference costs at the system level. Beyond cost efficiency, \sysname{} introduces distributed and disaggregated serving features, enabling scalable and flexible LLM inference across diverse workloads. Innovations such as the distributed KV cache pool, hybrid orchestration, and heterogeneous serving optimizer further push the boundaries of performance optimization, ensuring efficient resource utilization while maintaining low operational costs. These features collectively establish AIBrix as a highly adaptable and cost-effective solution for large-scale LLM deployment. By integrating deep inference engine optimizations with cloud-native infrastructure, \sysname{} provides a scalable, high-performance serving platform that bridges the gap between efficiency and flexibility in AI inference workloads.

\textbf{Limitations and future work.}
Some of our experiments do not fully evaluate routing strategies, heterogeneous serving under non-ideal workloads, limiting the ability to generalize these features across different workload characteristics. Additionally, profiling-based autoscaling and heterogeneous GPU scheduling currently rely on offline model profiling, which introduces an additional step that may not always be practical for dynamic workloads. To mitigate this, a potential solution is to streamline the profiling process by adopting roofline model analysis (\cite{imai2024predicting}), which can provide a more structured and lightweight approach to profiling heterogeneous inference performance.
Furthermore, we aim to expand evaluations to cover diverse real-world workload scenarios, refining routing strategies, GPU allocation mechanisms, and distributed orchestration to further enhance \sysname{}’s adaptability across various LLM deployment environments.  

Moving forward, we will continue to explore deeper co-design between inference engines and system architecture, ensuring that \sysname{} remains a highly optimized, scalable, and production-ready solution for LLM inference at scale.